\documentclass[twocolumn,aps,pra,showpacs]{revtex4}
\begin{document}
\title{Two-player quantum pseudotelepathy based on
recent all-versus-nothing violations of local realism}
\author{Ad\'{a}n Cabello}
\email{adan@us.es}
\affiliation{Departamento de F\'{\i}sica Aplicada II,
Universidad de Sevilla, 41012 Sevilla, Spain}
\date{\today}
%First version: 9 November 2005
%This version: 25 January 2006, after PRA's proofs

%%%%%%%%%%%%%%%%%%%%%%%%%%%%%%%%%%%%%%%%%%%%%%%%%%%%%%%%%%%%%%%%%%%

\begin{abstract}
We introduce two two-player quantum pseudotelepathy games based on
two recently proposed all-versus-nothing (AVN) proofs of Bell's
theorem [A. Cabello, Phys. Rev. Lett. {\bf 95}, 210401 (2005);
Phys. Rev. A {\bf 72}, 050101(R) (2005)]. These games prove that
Broadbent and M\'{e}thot's claim that these AVN proofs do not rule
out local-hidden-variable theories in which it is possible to
exchange unlimited information inside the same light cone
(quant-ph/0511047) is incorrect.
\end{abstract}

%%%%%%%%%%%%%%%%%%%%%%%%%%%%%%%%%%%%%%%%%%%%%%%%%%%%%%%%%%%%%%%%%%%

%Quantum optics
\maketitle

%%%%%%%%%%%%%%%%%%%%%%%%%%%%%%%%%%%%%%%%%%%%%%%%%%%%%%%%%%%%%%%%%%%

\section{Introduction}
\label{sec:1}

%%%%%%%%%%%%%%%%%%%%%%%%%%%%%%%%%%%%%%%%%%%%%%%%%%%%%%%%%%%%%%%%%%%

A game consisting of ``questions given to space-like separated
players who must give answers satisfying a certain relation with
the questions (\ldots), which cannot be won with certainty by
classical players who share common classical information, whereas
it can be won with certainty by quantum players who share
entanglement, is a pseudotelepathy game''~\cite{BM05}. Quantum
pseudotelepathy (QPT) is a particular type of
entanglement-assisted reduction of communication
complexity~\cite{CB97} in Yao's scenario~\cite{Yao79}. To my
knowledge, the first QPT game was the three-player game proposed
by Vaidman~\cite{Vaidman99,Vaidman01} from Mermin's
version~\cite{Mermin90} of the proof of Greenberger, Horne, and
Zeilinger~\cite{GHZ89} of Bell's theorem~\cite{Bell64}. A
rotationally invariant version of Vaidman's game was proposed
in~\cite{Cabello03}. Vaidman's game was extended for more players
in~\cite{BBT04b}.

Indeed, Vaidman~\cite{Vaidman01} was also the first to suggest
that a two-player QPT game could be derived from the two-observer
all-versus-nothing (AVN) proof proposed in~\cite{Cabello01}. QPT
games based on this AVN proof were developed by
Aravind~\cite{Aravind04}, Cleve and Mermin (see~\cite{Aravind04}),
and Brassard, Broadbent, and Tapp~\cite{BBT04a}. Indeed, the name
QPT was coined by Brassard, Broadbent, and
Tapp~\cite{BBT04a,BBT04b}.

Recently, two new two-observer AVN proofs have been
proposed~\cite{Cabello05a,Cabello05b}. In a recent
paper~\cite{BM05}, Broadbent and M\'{e}thot state that these new
AVN proofs: (a) do not ``rule out a certain class of of
local-hidden-variable models, those that do not use hidden
travelling information'' (i.e., those in which ``it is possible
that particles inside the same light cone can exchange unlimited
information''), and (b) cannot be translated into QPT games
involving two players.

Both claims are incorrect for the same reason. The two-observer
AVN proofs of Refs.~\cite{Cabello05a,Cabello05b} have two parts.
In the first one, the status of Einstein-Podolsky-Rosen's (EPR's)
element of reality~\cite{EPR35} is established for some local
observables. This attribution of elements of reality is based on
some predictions of quantum mechanics for a two-particle system
prepared in a specific quantum state. In the second part, a
logical inconsistency between these elements of reality appears
when some additional predictions of quantum mechanics are taken
into account. Broadbent and M\'{e}thot point out that the
predictions of the second part can be simulated within classical
physics. However, they ignore the predictions of the first part.
The point is that it is the whole set of quantum predictions which
{\em cannot} be simulated within classical physics.

This two-part structure is similar to that of Bell's original
proof~\cite{Bell64}. First, Bell points out that some predictions
of quantum mechanics for the singlet state $|\psi^{-}\rangle$
allow us to identify any local spin observable $X_j$ as an EPR
element of reality. Specifically, the quantum predictions he uses
are that $X_1 X_2 |\psi^{-}\rangle = -|\psi^{-}\rangle$ for any
$X$. He then derives an inequality which is valid under the
assumption that spin observables are elements of reality. In the
second part, he proves that another set of quantum predictions
violates this inequality. This second set of quantum predictions
is $\langle \psi^{-}|A_1 B_2|\psi^{-}\rangle = \langle
\psi^{-}|A_1 b_2 |\psi^{-}\rangle = \langle \psi^{-}|a_1 B_2
|\psi^{-}\rangle = - \langle \psi^{-}|a_1 b_2 |\psi^{-}\rangle =
1/\sqrt{2}$, for some specific $A$, $a$, $B$, and $b$ such that $A
\neq B$, $A \neq b$, $a \neq B$, and $a \neq b$~\cite{CHSH69}.
Note, however, that only the predictions of the second set are
tested in experimental violations of Bell's
inequalities~\cite{ADR82}.

Both Refs.~\cite{Cabello05a,Cabello05b} contain ``games,'' which
were {\em not} designed as QPT games, but (as it is specifically
stated) as tools ``to estimate the detection efficiency required
to avoid the detection loophole'' in Bell-type experiments based
on the proofs in Refs.~\cite{Cabello05a,Cabello05b}.

In this paper we show that the two-observer AVN proofs in
Refs.~\cite{Cabello05a,Cabello05b} can be translated into
two-player QPT games, without any mention of the concept of EPR's
element of reality. The fact that these games cannot always be won
using only classical resources proves indeed that no
local-hidden-variable theory (even those in which it is possible
to exchange unlimited information inside the same light cone) can
reproduce the predictions of quantum mechanics, and therefore
constitutes a refutation of both of Broadbent and M\'{e}thot's
claims.

%%%%%%%%%%%%%%%%%%%%%%%%%%%% Table I %%%%%%%%%%%%%%%%%%%%%%%%%%%%%%

\begin{table*}[bth]
\begin{center}
\begin{tabular}{c|cccccccc}
\hline \hline
Bob/Alice & $X_1 \& z_1?$ & $Z_1 \& x_1?$
& $Y_1 \& z_1?$ & $Z_1 \& y_1?$
& $X_1 \& x_1?$ & $X_1 \& y_1?$
& $Y_1 \& x_1?$ & $Y_1 \& y_1?$ \\
\hline
$X_2 \& z_2?$ & $\left\{\begin{array}{c} X_1 = X_2 z_2 \\ X_1 z_1 = X_2 \end{array}\right.$ & & $z_1 = z_2$ & & $X_1 = X_2 z_2$ & $X_1 = X_2 z_2$ & & \\
$Z_2 \& x_2?$ & & $\left\{\begin{array}{c} Z_1 x_1 = x_2 \\ x_1 = Z_2 x_2 \end{array}\right.$ & & $Z_1 = Z_2$ & $x_1 = Z_2 x_2$ & & $x_1 = Z_2 x_2$ & \\
$Y_2 \& z_2?$ & $z_1 = z_2$ & & $\left\{\begin{array}{c} Y_1 = -Y_2 z_2 \\ Y_1 z_1 = -Y_2 \end{array}\right.$ & & & & $Y_1 = -Y_2 z_2$ & $Y_1 = -Y_2 z_2$ \\
$Z_2 \& y_2?$ & & $Z_1 = Z_2$ & & $\left\{\begin{array}{c} Z_1 y_1 = -y_2 \\ y_1 = -Z_2 y_2 \end{array}\right.$ & & $y_1 = -Z_2 y_2$ & & $y_1 = -Z_2 y_2$ \\
$Y_2 \& y_2?$ & $\begin{array}{c} \mbox{} \\ \mbox{} \end{array}$ & & $Y_1 z_1 = -Y_2$ & $Z_1 y_1 = -y_2$ & $X_1 x_1 = Y_2 y_2$ & & & \\
$Y_2 \& x_2?$ & $\begin{array}{c} \mbox{} \\ \mbox{} \end{array}$ & $Z_1 x_1 = x_2$ & $Y_1 z_1 = -Y_2$ & & & $X_1 y_1 = Y_2 x_2$ & & \\
$X_2 \& y_2?$ & $X_1 z_1 = X_2$ & $\begin{array}{c} \mbox{} \\ \mbox{} \end{array}$ & & $Z_1 y_1 =-y_2$ & & & $Y_1 x_1 = X_2 y_2$ & \\
$X_2 \& x_2?$ & $X_1 z_1 = X_2$ & $Z_1 x_1 = x_2$ & $\begin{array}{c} \mbox{} \\ \mbox{} \end{array}$ & & & & & $Y_1 y_1 = X_2 x_2$ \\
\hline \hline
\end{tabular}
\end{center}
%\vspace{0.2cm}
\noindent TABLE I. {\small Rules of the two-player QPT game. Each
of two players Alice and Bob is asked one out of eight possible
questions. If the intersection between one of the questions Alice
is asked (upper row) and one of the questions Bob is asked (left
column) is empty, then it means that this particular combination
of questions never occurs in the game. For the other
28~combinations, the intersections show the requirements for
winning those particular rounds.}
\end{table*}

%%%%%%%%%%%%%%%%%%%%%%%%%%%%%%%%%%%%%%%%%%%%%%%%%%%%%%%%%%%%%%%%%%%

\section{Rules of the two-player QPT game}
\label{sec:2}

%%%%%%%%%%%%%%%%%%%%%%%%%%%%%%%%%%%%%%%%%%%%%%%%%%%%%%%%%%%%%%%%%%%

Consider a team of two players, Alice and Bob, each of them in a
spacelike separated region. Each of them is asked one out of eight
possible questions: (i)~what are $X$ and $z$?, (ii)~what are $Z$
and $x$?, (iii)~what are $Y$ and $z$?, (iv)~what are $Z$ and $y$?,
(v)~what are $X$ and $x$?, (vi)~what are $X$ and $y$?, (vii)~what
are $Y$ and $x$?, and (viii)~what are $Y$ and $y$? Each player
must give one of the following answers: $-1$ and $-1$, $-1$ and
$1$, $1$ and $-1$, or $1$ and $1$. The rules of the game can be
found in Table~I. If the intersection between one of the questions
Alice is asked and one of the questions Bob is asked is empty in
Table~I, then it means that this particular combination of
questions never occurs during the game. In the intersections for
the other 28~combinations, Table~I shows the requirements for
winning those particular rounds. For instance, if Alice is asked,
what are $X_1$ and $z_1$?, and Bob is asked, what are $X_2$ and
$z_2$?, they win if their answers satisfy $X_1 = X_2 z_2$ and $X_1
z_1 = X_2$.

%%%%%%%%%%%%%%%%%%%%%%%%%%%%%%%%%%%%%%%%%%%%%%%%%%%%%%%%%%%%%%%%%%%

\section{Proof that classical players cannot always win}
\label{sec:3}

%%%%%%%%%%%%%%%%%%%%%%%%%%%%%%%%%%%%%%%%%%%%%%%%%%%%%%%%%%%%%%%%%%%

Let us assume, as Broadbent and M\'{e}thot suggest, that Alice's
answer to $X_1$ can be different when $X_1$ is asked jointly with
$x_1$ than when $X_1$ is asked jointly with $y_1$ or when $X_1$ is
asked jointly with $z_1$. Let us denote these answers as
$v(X_1|x_1)$, $v(X_1|y_1)$, and $v(X_1|z_1)$, respectively. Using
a similar notation for the answers to all the other questions in
all possible scenarios, it can be immediately seen that, in order
to satisfy all the requirements given in Table~I, these answers
must satisfy the following 32~equations:
\begin{eqnarray}
v(X_1|z_1) & = & v(X_2|z_2) v(z_2|X_2), \label{e01} \\
v(X_1|z_1) v(z_1|X_1) & = & v(X_2|z_2), \label{e02} \\
v(z_1|Y_1) & = & v(z_2|X_2), \label{e11} \\
v(X_1|x_1) & = & v(X_2|z_2) v(z_2|X_2), \label{e21} \\
v(X_1|y_1) & = & v(X_2|z_2) v(z_2|X_2), \label{e24} \\
v(Z_1|x_1) v(x_1|Z_1) & = & v(x_2|Z_2), \label{e06} \\
v(x_1|Z_1) & = & v(Z_2|x_2) v(x_2|Z_2), \label{e07} \\
v(Z_1|y_1) & = & v(Z_2|x_2), \label{e16} \\
v(x_1|X_1) & = & v(Z_2|x_2) v(x_2|Z_2), \label{e22} \\
v(x_1|Y_1) & = & v(Z_2|x_2) v(x_2|Z_2), \label{e27} \\
v(z_1|X_1) & = & v(z_2|Y_2), \label{e03} \\
v(Y_1|z_1) & = & -v(Y_2|z_2) V(z_2|Y_2), \label{e12} \\
v(Y_1|z_1) v(z_1|Y_1) & = & -v(Y_2|z_2), \label{e13} \\
v(Y_1|x_1) & = & -v(Y_2|z_2) v(z_2|Y_2), \label{e28} \\
v(Y_1|y_1) & = & -v(Y_2|z_2) v(z_2|Y_2), \label{e30} \\
v(Z_1|x_1) & = & v(Z_2|y_2), \label{e08} \\
v(Z_1|y_1) v(y_1|Z_1) & = & -v(y_2|Z_2), \label{e17} \\
v(y_1|Z_1) & = & -v(Z_2|y_2) v(y_2|Z_2), \label{e18} \\
v(y_1|X_1) & = & -v(Z_2|y_2) v(y_2|Z_2), \label{e25} \\
v(y_1|Y_1) & = & -v(Z_2|y_2) v(y_2|Z_2), \label{e31} \\
v(Y_1|z_1) v(z_1|Y_1) & = & -v(Y_2|y_2), \label{e14} \\
v(Z_1|y_1) v(y_1|Z_1) & = & -v(y_2|Y_2), \label{e19} \\
v(X_1|x_1) v(x_1|X_1) & = & v(Y_2|y_2) v(y_2|Y_2), \label{e23} \\
v(Z_1|x_1) v(x_1|Z_1) & = & v(x_2|Y_2), \label{e09} \\
v(Y_1|z_1) v(z_1|Y_1) & = & -v(Y_2|x_2), \label{e15} \\
v(X_1|y_1) v(y_1|X_1) & = & v(Y_2|x_2) v(x_2|Y_2), \label{e26} \\
v(X_1|z_1) v(z_1|X_1) & = & v(X_2|y_2), \label{e04} \\
v(Z_1|y_1) v(y_1|Z_1) & = & -v(y_2|X_2), \label{e20} \\
v(Y_1|x_1) v(x_1|Y_1) & = & v(X_2|y_2) v(y_2|X_2), \label{e29} \\
v(X_1|z_1) v(z_1|X_1) & = & v(X_2|x_2), \label{e05} \\
v(Z_1|x_1) v(x_1|Z_1) & = & v(x_2|X_2), \label{e10} \\
v(Y_1|y_1) v(y_1|Y_1) & = & v(X_2|x_2) v(x_2|X_2). \label{e32}
\end{eqnarray}

In order to satisfy Eqs.~(\ref{e01}), (\ref{e21}), and
(\ref{e24}),
\begin{equation}
v(X_1|z_1) = v(X_1|x_1) = v(X_1|y_1) \equiv v(X_1).
\end{equation}
In order to satisfy Eqs.~(\ref{e07}), (\ref{e22}), and
(\ref{e27}),
\begin{equation}
v(x_1|Z_1) = v(x_1|X_1) = v(x_1|Y_1) \equiv v(x_1).
\end{equation}
In order to satisfy Eqs.~(\ref{e12}), (\ref{e28}), and
(\ref{e30}),
\begin{equation}
v(Y_1|z_1) = v(Y_1|x_1) = v(Y_1|y_1) \equiv v(Y_1).
\end{equation}
In order to satisfy Eqs.~(\ref{e18})--(\ref{e31}),
\begin{equation}
v(y_1|Z_1) = v(y_1|X_1) = v(y_1|Y_1) \equiv v(y_1).
\end{equation}
Multiplying Eqs.~(\ref{e06}) and (\ref{e07}) and taking into
account Eq.~(\ref{e16}), we find that
\begin{equation}
v(Z_1|x_1) = v(Z_1|y_1) \equiv v(Z_1).
\end{equation}
Multiplying Eqs.~(\ref{e01}) and (\ref{e02}) and taking into
account Eq.~(\ref{e11}), we find that
\begin{equation}
v(z_1|X_1) = v(z_1|Y_1) \equiv v(z_1).
\end{equation}
In order to satisfy Eqs.~(\ref{e02}), (\ref{e04}), and
(\ref{e05}),
\begin{equation}
v(X_2|z_2) = v(X_2|y_2) = v(X_2|x_2) \equiv v(X_2).
\end{equation}
In order to satisfy Eqs.~(\ref{e06}), (\ref{e09}), and
(\ref{e10}),
\begin{equation}
v(x_2|Z_2) = v(x_2|Y_2) = v(x_2|X_2) \equiv v(x_2).
\end{equation}
In order to satisfy Eqs.~(\ref{e13}), (\ref{e14}), and
(\ref{e15}),
\begin{equation}
v(Y_2|z_2) = v(Y_2|y_2) = v(Y_2|x_2) \equiv v(Y_2).
\end{equation}
In order to satisfy Eqs.~(\ref{e17}), (\ref{e19}), and
(\ref{e20}),
\begin{equation}
v(y_2|Z_2) = v(y_2|Y_2) = v(y_2|X_2) \equiv v(y_2).
\end{equation}
Multiplying Eqs.~(\ref{e06}) and (\ref{e07}) and taking into
account Eq.~(\ref{e08}), we find that
\begin{equation}
v(Z_2|x_2) = v(Z_2|y_2) \equiv v(Z_2).
\end{equation}
Multiplying Eqs.~(\ref{e01}) and (\ref{e02}) and taking into
account Eq.~(\ref{e03}), we find that
\begin{equation}
v(z_2|X_2) = v(z_2|Y_2) \equiv v(z_2).
\end{equation}
Therefore, we have established that, in order to win some of the
rounds, Alice's answer to $X_1$ must be the same when $X_1$ is
asked jointly with $x_1$, when $X_1$ is asked jointly with $y_1$,
and when $X_1$ is asked jointly with $z_1$. Analogously for
Alice's answers to $x_1$, $Y_1$, $y_1$, $Z_1$, and $z_1$ and Bob's
answers to $X_2$, $x_2$, $Y_2$, $y_2$, $Z_2$, and
$z_2$~\cite{comment}.

Having proven this, there are many ways to prove that the players
cannot win all rounds (see~\cite{Cabello05b} for details). For
instance, in order to satisfy also Eqs.~(\ref{e21}), (\ref{e28}),
(\ref{e23}), and (\ref{e29}),
\begin{eqnarray} v(X_1) & = & v(X_2) v(z_2),
\label{valuno} \\
v(Y_1) & = & -v(Y_2) v(z_2),
\label{valcinco} \\
v(X_1) v(x_1) & = & v(Y_2) v(y_2),
\label{valnueve} \\
v(Y_1) v(x_1) & = & v(X_2) v(y_2),
\label{valonce}
\end{eqnarray}
respectively. However, it is impossible to assign the values $-1$
or $1$ in a way consistent with all
Eqs.~(\ref{valuno})--(\ref{valonce}) since the product of
Eqs.~(\ref{valuno})--(\ref{valonce}) gives $1=-1$. We therefore
conclude that the players cannot always win the game.

%%%%%%%%%%%%%%%%%%%%%%%%%%%%%%%%%%%%%%%%%%%%%%%%%%%%%%%%%%%%%%%%%%%

\section{Entanglement-assisted strategy}
\label{sec:4}

%%%%%%%%%%%%%%%%%%%%%%%%%%%%%%%%%%%%%%%%%%%%%%%%%%%%%%%%%%%%%%%%%%%

There is, however, a quantum entanglement-assisted strategy that
allows the players to always win the game. Suppose that Alice and Bob
share two photons entangled both in polarization and in path
degrees of freedom prepared in the state
\begin{eqnarray}
|\psi\rangle & = & \frac{1}{2} ( |H u\rangle_1 |H u\rangle_2 + |H
d\rangle_1 |H d\rangle_2 + |V u\rangle_1 |V u\rangle_2 \nonumber \\
& & - |V d\rangle_1 |V d\rangle_2), \label{benasque}
\end{eqnarray}
where $|H\rangle_j$ and $|V\rangle_j$ represent horizontal and
vertical polarization and $|u\rangle_j$ and $|d\rangle_j$ denote
two orthonormal path states for photon $j$. Consider also six
local observables on photon $j$: three for polarization degrees of
freedom, defined by the operators
\begin{eqnarray}
X_j & = & |H\rangle_j \langle V|+|V\rangle_j \langle H|, \\
Y_j & = & i \left(|V\rangle_j \langle H|-|H\rangle_j \langle V|\right), \\
Z_j & = & |H\rangle_j \langle H|-|V\rangle_j \langle V|,
\end{eqnarray}
and three for path degrees of freedom,
\begin{eqnarray}
x_j & = & |u\rangle_j \langle d|+|d\rangle_j \langle u|, \\
y_j & = & i \left(|d\rangle_j \langle u|-|u\rangle_j \langle
d|\right),
\\
z_j & = & |u\rangle_j \langle u|-|d\rangle_j \langle d|.
\end{eqnarray}
Each of these observables can take two values: $-1$ or $1$.

The state~(\ref{benasque}) satisfies the following equations:
\begin{eqnarray}
Z_1 Z_2 |\psi \rangle & = & |\psi \rangle, \label{eqmenosuno} \\
z_1 z_2 |\psi \rangle & = & |\psi \rangle, \label{eqcero} \\
X_1 X_2 z_2 |\psi \rangle & = & |\psi \rangle, \label{equno} \\
x_1 Z_2 x_2 |\psi \rangle & = & |\psi \rangle, \label{eqdos} \\
X_1 z_1 X_2 |\psi \rangle & = & |\psi \rangle, \label{eqtres} \\
Z_1 x_1 x_2 |\psi \rangle & = & |\psi \rangle, \label{eqcuatro} \\
Y_1 Y_2 z_2 |\psi \rangle & = & -|\psi \rangle, \label{eqcinco} \\
y_1 Z_2 y_2 |\psi \rangle & = & -|\psi \rangle, \label{eqseis} \\
Y_1 z_1 Y_2 |\psi \rangle & = & -|\psi \rangle, \label{eqsiete} \\
Z_1 y_1 y_2 |\psi \rangle & = & -|\psi \rangle, \label{eqocho} \\
X_1 x_1 Y_2 y_2 |\psi \rangle & = & |\psi \rangle, \label{eqnueve} \\
X_1 y_1 Y_2 x_2 |\psi \rangle & = & |\psi \rangle, \label{eqdiez} \\
Y_1 x_1 X_2 y_2 |\psi \rangle & = & |\psi \rangle, \label{eqonce} \\
Y_1 y_1 X_2 x_2 |\psi \rangle & = & |\psi \rangle. \label{eqdoce}
\end{eqnarray}
Therefore, if the players give as answers the results of the
corresponding measurements on their photons, then these answers
satisfy {\em all} Eqs.~(\ref{e01})--(\ref{e32}). This result,
together with the result proved in Sec.~\ref{sec:3}, shows that
the game presented in Sec.~\ref{sec:2} is a QPT game according to
the definition given in Sec.~\ref{sec:1}.

%%%%%%%%%%%%%%%%%%%%%%%%%%%%%%%%%%%%%%%%%%%%%%%%%%%%%%%%%%%%%%%%%%%

\section{Optimal classical strategy}
\label{sec:5}

%%%%%%%%%%%%%%%%%%%%%%%%%%%%%%%%%%%%%%%%%%%%%%%%%%%%%%%%%%%%%%%%%%%

The choice of the best classical strategy depends on the relative
frequency of the possible combinations of questions. Assuming that
the 28~possible combinations occur with the same frequency, if the
players always answer $1$ to any question except to $Y_2$ and
$y_2$, for which Bob gives the answer $-1$, then they win in 26 of
the 28~combinations; this strategy fails to satisfy
Eqs.~({\ref{e26}) and ({\ref{e29}). As a careful examination
reveals, this classical strategy is optimal.

%%%%%%%%%%%%%%%%%%%%%%%%%%%% Table II %%%%%%%%%%%%%%%%%%%%%%%%%%%%%%

\begin{table*}[tbh]
\begin{center}
\begin{tabular}{c|ccccc}
\hline \hline Bob/Alice & $X_1 \& z_1?$ & $Y_1 \& z_1?$ & $Z_1 \& y_1?$ & $X_1 \& x_1?$ & $Y_1 \& x_1?$ \\
\hline
$X_2 \& z_2?$ & $\left\{\begin{array}{c} X_1 = X_2 z_2 \\ X_1 z_1 = X_2 \end{array}\right.$ & $z_1=z_2$ & & $X_1 = X_2 z_2$ & \\
$Z_2 \& x_2?$ & $\begin{array}{c} \mbox{} \\ \mbox{} \end{array}$ & & & $x_1 = Z_2 x_2$ & $x_1 = Z_2 x_2$ \\
$Y_2 \& z_2?$ & $z_1=z_2$ & $\left\{\begin{array}{c} Y_1 = -Y_2 z_2 \\ Y_1 z_1 = -Y_2 \end{array}\right.$ & & & $Y_1 = -Y_2 z_2$ \\
$Y_2 \& y_2?$ & $\begin{array}{c} \mbox{} \\ \mbox{} \end{array}$ & $Y_1 z_1 = -Y_2$ & $Z_1 y_1 = -y_2$ & $X_1 x_1 = Y_2 y_2$ & \\
$X_2 \& y_2?$ & $X_1 z_1 = X_2$ & $\begin{array}{c} \mbox{} \\ \mbox{} \end{array}$ & $Z_1 y_1 = -y_2$ & & $Y_1 x_1 = X_2 y_2$ \\
\hline \hline
\end{tabular}
\end{center}
%\vspace{0.2cm}
\noindent TABLE II. {\small Rules of the simpler two-player QPT
game. Each of two players Alice and Bob is asked one out of five
possible questions. The intersections show the requirements for
winning those particular rounds. An empty intersection means that
this particular combination does no occur.}
\end{table*}

%%%%%%%%%%%%%%%%%%%%%%%%%%%%%%%%%%%%%%%%%%%%%%%%%%%%%%%%%%%%%%%%%%%

\section{Simpler two-player QPT game}
\label{sec:6}

%%%%%%%%%%%%%%%%%%%%%%%%%%%%%%%%%%%%%%%%%%%%%%%%%%%%%%%%%%%%%%%%%%%

The QPT game described in the previous sections has the virtue of
containing not only the simpler two-observer AVN proof presented
in Ref.~\cite{Cabello05a}, but also the extended AVN proof of
Ref.~\cite{Cabello05b}. However, a simpler QPT game can be
obtained from the simpler two-observer AVN proof of
Ref.~\cite{Cabello05a}. The rules of this new game are indeed a
subset of the rules of the previous game. The rules of the new
game are explained in Table~II.

The proof that classical players cannot always win this game
follows from the fact that, in order to satisfy all the
requirements given in Table~II, Eqs.~(\ref{e01})--(\ref{e21}),
(\ref{e22})--(\ref{e28}), (\ref{e14})--(\ref{e23}), and
(\ref{e04})--(\ref{e29}) must be satisfied.

From Eqs.~(\ref{e01}) and (\ref{e21}) it follows that $v(X_1|z_1)=
v(X_1|x_1) \equiv v(X_1)$. From Eqs.~(\ref{e12}) and (\ref{e28})
it follows that $v(Y_1|z_1)= v(Y_1|x_1) \equiv v(Y_1)$. From
Eqs.~(\ref{e22}) and (\ref{e27}) it follows that $v(x_1|X_1)=
v(x_1|Y_1) \equiv v(x_1)$. From Eqs.~(\ref{e02}) and (\ref{e04})
it follows that $v(X_2|z_2)= v(X_2|y_2) \equiv v(X_2)$. From
Eqs.~(\ref{e13}) and (\ref{e14}) it follows that $v(Y_2|z_2)=
v(Y_2|y_2) \equiv v(Y_2)$. From Eqs.~(\ref{e19}) and (\ref{e20})
it follows that $v(y_2|Y_2)= v(y_2|X_2) \equiv v(y_2)$. From
Eq.~(\ref{e11}) and the product of Eqs.~(\ref{e12}) and
(\ref{e13}) [or, alternatively, from Eq.~(\ref{e03}) and the
product of Eqs.~(\ref{e01}) and (\ref{e02})] it follows that
$v(z_2|X_2) = v(z_2|Y_2) \equiv v(z_2)$~\cite{comment}.

In order to satisfy also Eqs.~(\ref{e21}), (\ref{e28}),
(\ref{e23}), and (\ref{e29}), Eqs.~(\ref{valuno})--(\ref{valonce})
must be satisfied. However, as stated before, it is impossible to
assign the values $-1$ or $1$ in a way consistent with all
Eqs.~(\ref{valuno})--(\ref{valonce}) since the product of them
gives $1=-1$. We therefore conclude that the players cannot always
win this new game. However, the same entanglement-assisted
strategy described in Sec.~\ref{sec:4} would allow them to always
win the game. Therefore, the game in Table~II is also a QPT game.

Assuming that the 14~possible combinations of questions occur with
the same frequency, an optimal classical strategy (for instance,
all answers are $1$, except $Y_2$ and $y_2$ which are $-1$) allows
the players to win with probability~$13/14$ [it fails to satisfy
Eq.~(\ref{e29}); other frequencies of the combinations give lower
probabilities].

%%%%%%%%%%%%%%%%%%%%%%%%%%%%%%%%%%%%%%%%%%%%%%%%%%%%%%%%%%%%%%%%%%%

\section{Conclusions}
\label{sec:7}

%%%%%%%%%%%%%%%%%%%%%%%%%%%%%%%%%%%%%%%%%%%%%%%%%%%%%%%%%%%%%%%%%%%

A simple four-player QPT game can be derived from
Eqs.~(\ref{valuno})--(\ref{valonce}) and the four-qubit version of
the state~(\ref{benasque}). The interesting point of the two QPT
games presented in this paper is that they are two-player QPT
games. Both have been derived from the two-observer AVN proofs of
Refs.~\cite{Cabello05a,Cabello05b}. The main difference with
respect to Refs.~\cite{Cabello05a,Cabello05b} is that here we have
not explicitly used the EPR criterion for elements of reality.
Therefore, these QPT games prove that, contrary to Broadbent and
M\'{e}thot's claim, the AVN proofs of
Refs.~\cite{Cabello05a,Cabello05b} rule out all
local-hidden-variable theories, even those in which it is possible
to exchange unlimited information inside the same light cone.
Besides the challenge of presenting new two-player QPT games, the
main goal of this paper is to dispel any possible doubts about the
correctness of these AVN proofs and of the works in progress based
on them.

%%%%%%%%%%%%%%%%%%%%%%%%%%%%%%%%%%%%%%%%%%%%%%%%%%%%%%%%%%%%%%%%%%%

\section*{Acknowledgments}

%%%%%%%%%%%%%%%%%%%%%%%%%%%%%%%%%%%%%%%%%%%%%%%%%%%%%%%%%%%%%%%%%%%

The author thanks Anne Broadbent for stimulating discussions and
acknowledges support by Projects Nos.~FIS2005-07689 and~FQM-239.

%%%%%%%%%%%%%%%%%%%%%%%%%%% References %%%%%%%%%%%%%%%%%%%%%%%%%%%%

%%%%%%%%%%%%%%%%%%%%%%%%%%%%%%%%%%%%%%%%%%%%%%%%%%%%%%%%%%%%%%%%%%%

\end{document}